\documentclass[aps,prl,twocolumn,showpacs,superscriptaddress,groupedaddress]{revtex4}  
\usepackage{graphicx}
\usepackage{hyperref}
\begin{document}
\title{Search for Neutron Flux Generation in a Plasma Discharge Electrolytic Cell}
\author{R.~Faccini, A. Pilloni, A.D.~Polosa}
 \affiliation{Dipartimento di Fisica, Sapienza Universit\`a di Roma and INFN Sezione di Roma,\\ Piazzale Aldo Moro 2, I-00185 Roma, Italy}
\author{ M. Angelone, E. Castagna, S. Lecci, A. Pietropaolo, M. Pillon, M. Sansovini, F. Sarto, V.~Violante}
 \affiliation{Enea, Centro Ricerche di Frascati, Via E. Fermi 45, I-00044 Frascati, Italy }
 \author{R. Bedogni, A. Esposito}
 \affiliation{INFN Laboratori Nazionali di Frascati, Via E. Fermi 40, I-00044 Frascati, Italy }


\thispagestyle{empty}

\begin{abstract}
Following some recent unexpected hints of neutron production in setups like high-voltage atmospheric discharges and plasma discharges in electrolytic cells, we present a measurement of the neutron flux in a configuration similar to the latter.
We use two different types of neutron detectors, poly-allyl-diglicol-carbonate (PADC, aka CR-39) tracers 
and  Indium disks. At 95\% C.L. we provide an upper limit of 1.5 neutrons~cm$^{-2}$~s$^{-1}$ for the thermal neutron flux at 
$\approx 5$ cm from the center of the cell. Allowing for a higher energy neutron component the largest allowed flux is 64 neutrons~cm$^{-2}$~s$^{-1}$. This upper limit is two orders of magnitude smaller than what previously claimed in an electrolytic cell plasma discharge experiment.
Furthermore the behavior of the CR-39 is discussed  to point our possible sources of spurious signals.

\end{abstract}
 \maketitle


\label{sec:intro}
The possibility that completely new sources of neutrons could be devised and exploited is a topic of paramount importance on both  the applicative and theoretical grounds. Some impressive hints on neutron production in high-voltage atmospheric discharges are reported in a recent paper~\cite{fulmini} as well as production of neutrons in plasma discharges in electrolytic cells are discussed in~\cite{Cirillo}. Great resonance had also the approval of the patent~\cite{patent} similarly claiming the existence of a neutron producing  electrolytic cell. 

We focus here on  an experimental set-up similar to the one discussed in~\cite{Cirillo, patent} albeit the underlying physical mechanisms could be in common with Refs.~\cite{fulmini}, the so called Low Energy Nuclear Reactions (LENR).

From the theoretical point of view, there is an open debate on whether inverse-$\beta$ nuclear transmutations could justify such observations~\cite{Widom, Ciuchi}.
Experiments are therefore needed to guide the research in the field.
To this aim,
we have  reproduced  the experimental setup of Ref.~\cite{Cirillo} and searched for a neutron flux with two different detection techniques: the CR-39 detectors, also used in the experiment on atmospheric discharges~\cite{fulmini} and Indium activation, commonly used in neutron flux measurements.  A dedicated effort was devoted to understanding  the behavior of the CR-39 detectors.

\begin{figure}[b!]
\centering
\includegraphics[width=\columnwidth]{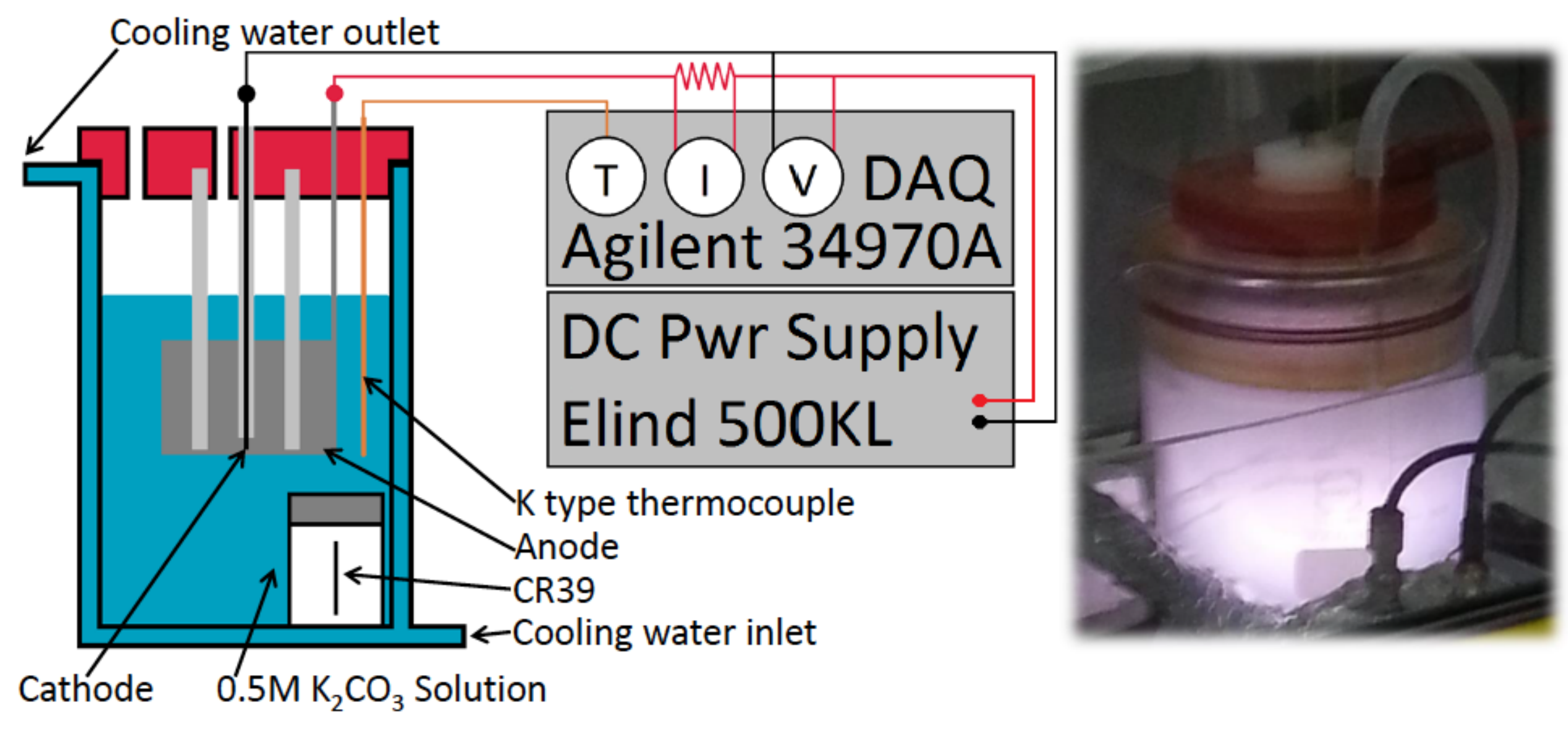}
\caption{Scheme of the experimental setup. At the bottom right a picture of the cell with plasma.}
\label{fig:setup}
\end{figure}

{\bf{The experimental setup.}}
The two main elements of this measurement are the electrolytic cell and the neutron detectors.
For reasons that will be evident after the results are presented,  the experiment consisted of three data-taking campaigns, Run1, Run2 and Run3, that took place  in February-March, June, and September 2013, respectively.
The experimental setup and the changes between the campaigns will be discussed in the following.

{\it{The electrolytic cell.}}
Plasma electrolysis can be considered an hybrid between the  conventional electrolysis process and the atmospheric-pressure plasma production.  In aqueous solutions the electrochemical process can produce plasma with applied voltages of at least $\approx 130$~V. 

The electrochemical cell we use (see Fig.~\ref{fig:setup})   consists of a water cooled  pyrex vessel 
with the anodic and the cathodic areas separated by a quartz cylinder of 45~mm inner diameter, 4~mm thick. The anode consists of a cylindric grid of titanium plated with platinum positioned outside the quartz cylinder, while the tungsten cathode is a  pipe having outer diameter of 2~mm and lateral wall thickness of 1.4~mm during Run1 or a cylindrical rod with a diameter of 2.1 mm during Run2 and Run3. The cathode is partially covered by a quartz tube, in order to reduce its conductive active area, thus increasing the electric impedance of the electrochemical system. This makes the plasma discharge rise at the electrolyte-cathode interface. The electrolyte consists of 1.16 liters of  0.5M K$_2$CO$_3$ aqueous solution.

The cell is covered with an araldite cap having two passages, through which the gases produced both in the cathodic zone (hydrogen and oxygen basically produced by plasma discharge) and in the anodic zone (oxygen only from the Faradic current) are separately collected. We use a bubble flowmeter to measure the anodic gas flux and from it we derive the Faradic current. 
The experiment has been performed in voltage-control conditions. The power supplier is an Elind model 500KL2,4/6. The acquisition system is composed by an Agilent 34970A data logger and by a National Instruments NIDAQPad-6015. Electrolyte temperature has been measured by a K type thermocouple.

Plasma formation has been observed (see Fig.~\ref{fig:setup})  within a voltage range from 150~V to 300~V, with current ranging from 1.5~A to 3.0~A. The average oxygen flow value was measured to be  $1.5 \times 10^{-2}$~cm$^3$~s$^{-1}$,
 corresponding to a  Faradic current of 0.25~A. The exceeding current is responsible for  plasma discharge formation.  

{\it{The neutron detectors.}}
To detect the  neutron flux, two different types of detectors were used:  poly-allyl-diglicol-carbonate (PADC, CR-39 in the following)  track detectors 
and Indium disks.  To optimize their sensitivity they were located  close to the cathode, where the plasma is produced, i.e. 5 cm when placed outside the cell, 2 cm when inside.

{\it{The CR-39 detectors.}}
The CR-39 detectors produced by Intercast used in this tests are intrinsically sensitive to fast neutrons that produce tracks on them.  Thermal neutrons can also be detected by  wrapping the detectors in Boron: the $^{10}B(n,\alpha)^7Li$ reaction
generate $\alpha$ particles of $<1.8$MeV that create tracks in CR-39.

The samples used during the Run1 campaign belonged to a different stock than those used during the Run2 and Run3 campaigns. Each stock is expected to have different sensitivities and background levels and will be considered separately in the following.
The CR-39 detectors were coupled with a  thin (50$\mu$m) layer of pure $^{10}$B .
Furthermore, since in the original experiment~\cite{Cirillo} the CR-39 detectors were completely covered by a thick layer of boric acid, we also used some of the Run1 detectors with a $\approx 1$ cm thick surrounding layer of Boron.

To estimate the number of neutrons that traversed the detector since its production, the CR-39 slabs are etched for 90 min in a 6.25 N KOH solution at 70$^o$C. The number of tracks consistent with $\alpha$ particles is then estimated with an in-house automated reader consisting of an epi-illumination
microscope on which a 8 Mpixel CCD camera is mounted~\cite{bedogni}.
 Such number, divided by the detector area (2.224~cm$^2$), yields the track density ($D$).

\begin{figure}[tbh!]
\centering
\includegraphics[scale=0.38]{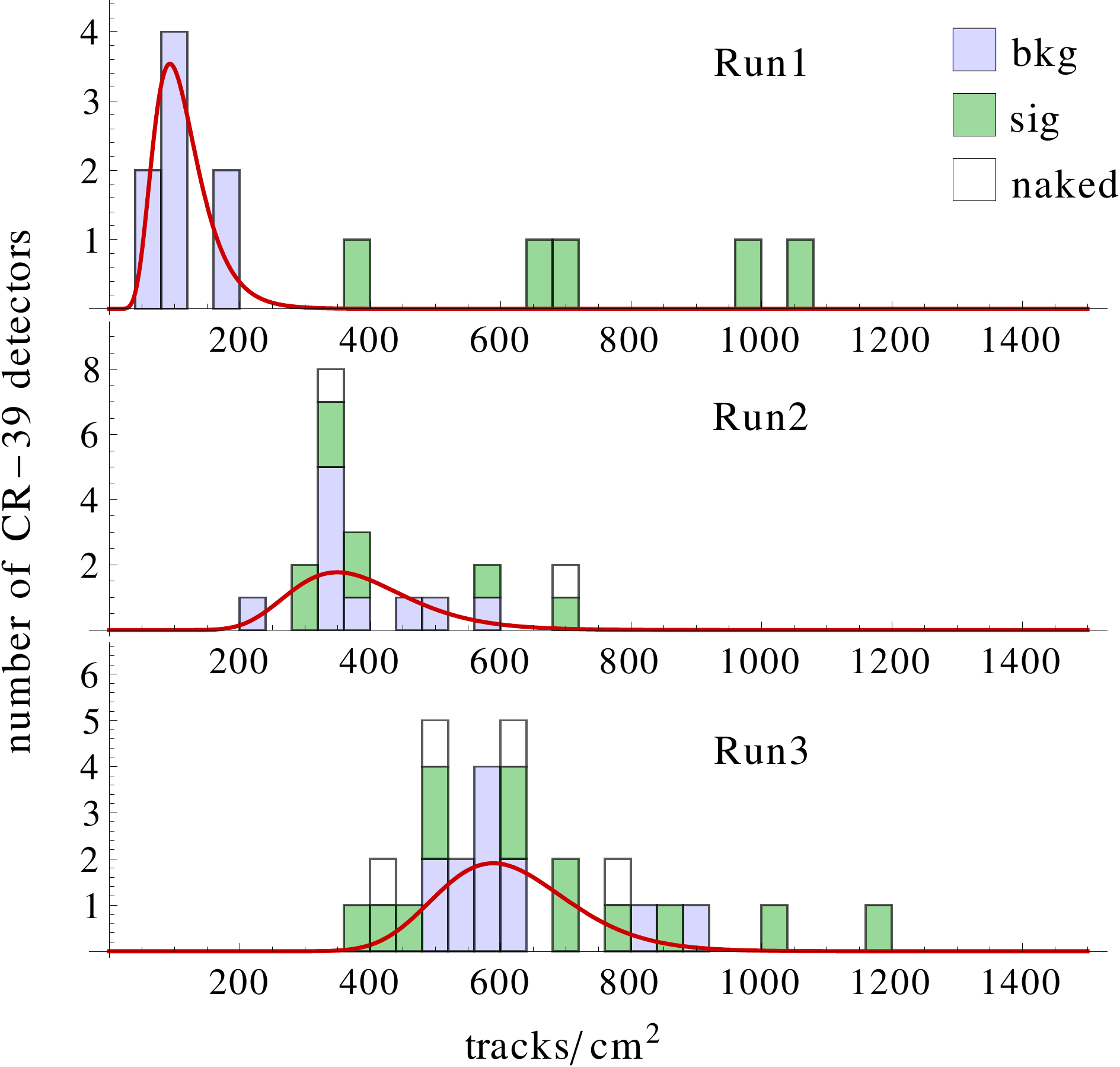}
\caption{Distribution of the track density $D$ in background detectors (pale blue histograms), irradiated detectors (dark green histograms) and detectors not  coupled with boron (open histograms). The data from the three data-taking campaigns are separated. The lines represent the results of  fits of the background distributions to a log-normal function.}
\label{fig:results}
\end{figure}

Spurious signals or ambient backgrounds are subtracted by analyzing a set of detectors not exposed to the neutron source under study. The distributions of $D$ in the three background samples used in this study and described later are shown in Fig.~\ref{fig:results}. Indicating with  $D_{bkg}$ the mean value of $D$  on such detectors we estimate the neutron flux on the irradiated detectors as
$F=\left(D-D_{bkg}\right)\!\big/\left(cT\right)$ where $T$ is the duration of the exposure. The calibration constant $c$ was obtained separately for the two stocks of detectors by exposing part of the detectors to a  calibrated thermal neutron flux of $(1.20\pm0.06)\!\times\!10^4\, \text{cm}^{-2}\text{s}^{-1}$ at the Casaccia ENEA-INMRI facility for different exposure times.

There are four systematic error contributions to such measurements: the Poisson fluctuations of the number of observed tracks, a 10\% uncertainty on the number of tracks due to the  counting process, the uncertainty on the calibration $c$, and the propagation of the error on $D_{bkg}$.
When quoting the measured fluxes all errors will be summed in quadrature.

We also estimated the compatibility of irradiated samples with the background: we first fit the measured $D$ distributions on background with a log-normal distribution, $L\left(D\,|\,\mu,\sigma\right)=\exp\left[-\left(\log D-\mu\right)^2 \!\big/ \left(2\sigma^2\right)\right]\Big/\left(\sqrt{2\pi}x\sigma\right)$, with $D>0$~\cite{bedogni2} and then we estimate the probability of a measured value of $D$ to be consistent with the background to be ${\cal{P}}(D)=\int_0^{D}L\left(x\,|\,\mu,\sigma\right) dx$.

To quantify the sensitivity of a negative result, when ${\cal{P}}<99\%$ we also compute, with a bayesian approach, the minimum neutron flux  that is inconsistent with the measured track density at 95\% C.L., $F_{lim}$. 

{\it{The Indium disks.}}
If exposed to a neutron flux the Indium produces, by the $^{115}$In(n,$\gamma$)$^{116}$In transition,  mainly  1293, 1097, and 416  keV $\gamma$ lines with a half life of 54 minutes. 
The photon production cross-section depends on the energy of the neutrons, ranging from $\approx$1.5 barn for the expected spectrum from partially moderated fast neutrons ("PM" in the following) to $\approx 60$ barn for thermal neutrons.

Indium disks of a diameter of 5 cm and 30 g total mass are therefore used as neutron detectors with a two stages process: they are first exposed to a potential neutron flux and then their activity is measured by means of a high purity absolutely calibrated germanium detector of 60\% efficiency. In absence of prior knowledge of the neutron spectrum, the conversion between the measured activity and the neutron flux is established by exposing the disks to a moderated AmBe neutron source of known rate ($\approx 2\times 10^6$n/s ) and by correcting for the decay times of the Indium isotopes.
 In case no signal is observed, the 95\% C.L. upper limit on the photon rate, and therefore on the neutron flux, is computed from the background level, the germanium detector efficiency and the conversion factors. 

As a check, an Indium disk and a CR-39 detector were also simultaneously exposed  to the neutrons produced by a $5.5\times 10^4$ n/s AmB neutron source placed at the cathode of the electrolytic cell filled with water and not powered. The comparison with detailed simulations confirmed the calibration of the Indium technique, while the consistency between Indium disks and CR-39 detectors within 30\% showed that within this range the calibration of the CR-39 detectors holds up to PM neutrons.

We will report upper limits computed assuming a thermal neutron flux. In case of a PM neutron spectrum  the limits would be a factor $\approx$40 larger.

\begin{table}[b!]
\begin{center}
\begin{tabular}{|l || c |c || c | c||}
\hline
Run 	& Duration	& Voltage Range 	& n-flux (In)	 & n-flux  (CR-39)  \\
       	& [']        		& 	[V]	  		& [n cm$^{-2}$ s$^{-1}$]   	&  [n cm$^{-2}$ s$^{-1}$]  \\ \hline 
1A     & 8              	& 150-200			&			 & $280 \pm 32$  (ins.) \\
 & & & & $259 \pm 30$  (ins.) \\
1B     & 5              	& 250 			&$<1.5$ & $275 \pm 35$\\
1C    & 12              	& 150-200			& $<0.7$   & $109\pm14$  (ins.)\\	
1D     & 4              	& 150-300 &		& 			 $166\pm32$  (ins.)\\ 
\hline
2     & 13             	& 220-300  		&	$<0.6$	& $< 49$, $< 14$ \\
 & & & & $< 23$, $< 18$ (Al)\\
 & & & & $< 16$, $< 21$ (Cd)\\
\hline
3A     & 20              	& 150 		&	$<0.4$	&  $< 26$, $< 27$\\
  & & & & $89 \pm 15$ (Al)\\
  & & & & $< 19$, $< 5$ (Cd)\\
3B     & 21              	& 200-300 		&	$<0.4$	& $< 11$, $39 \pm 12$\\
  & & & & $< 39$, $< 8$ (Cd)\\
  & & & & $56 \pm 13$ (Al)\\
\hline

	 \end{tabular}
\end{center}
\caption{\small{Characteristics of the runs and results. CR-39 detectors marked with (ins.) were placed inside the cells; the ones marked with (Al) and (Cd) were wrapped in aluminum and cadmium. Upper limits are at 95\% C.L.  The Indium upper limits are computed assuming a thermal neutron flux. For a PM neutron spectrum the limits are a factor $\approx$40 larger.}}
\label{tab:results}
\end{table}

{\bf{Results of the Run1 data taking.}}
\label{sec:results}
In our first experimental campaign the CR-39 detectors were calibrated by exposing three sets of three detectors  at the ENEA-INMRI thermal neutron  facility  for 1, 2, and 5 minutes respectively.
 The background was estimated from a sample of eight detectors that were not irradiated and not wrapped with boron and that were analyzed at the same time as the calibration samples. From a linear fit  to  the dependence of the measured tracks density after background subtraction on  the known neutron flux, the calibration constant is estimated to be $c=(6.9\pm0.3)\times 10^{-3}$ tracks/neutrons. 

At the same time, CR-39 detectors covered with a thick boron coating were exposed to the calibration flux, but no significant signal was observed: the Boron indeed absorbs the neutrons and stops the $\alpha$ particles. This observation contrasts with the measurement of neutron fluxes made in a similar configuration in Ref.~\cite{Cirillo}.

Several weeks afterwards we performed four different runs where we placed neutron detectors both close  to the electrolytic cell and inside it. The runs differed by voltage applied to the cell,  by duration, and by type and number of CR-39 detectors located inside and outside the cell.  During two of these runs we also used the Indium disks. The results are summarized in Tab.~\ref{tab:results}, where a letter is appended to Run1 to distinguish among the configurations. In analyzing the results we used the same background estimated for calibration, which means that the exposed detectors were analyzed 
40 days after the corresponding background ones.

\begin{figure}[tbh!]
\centering
\includegraphics[width=0.7\columnwidth]{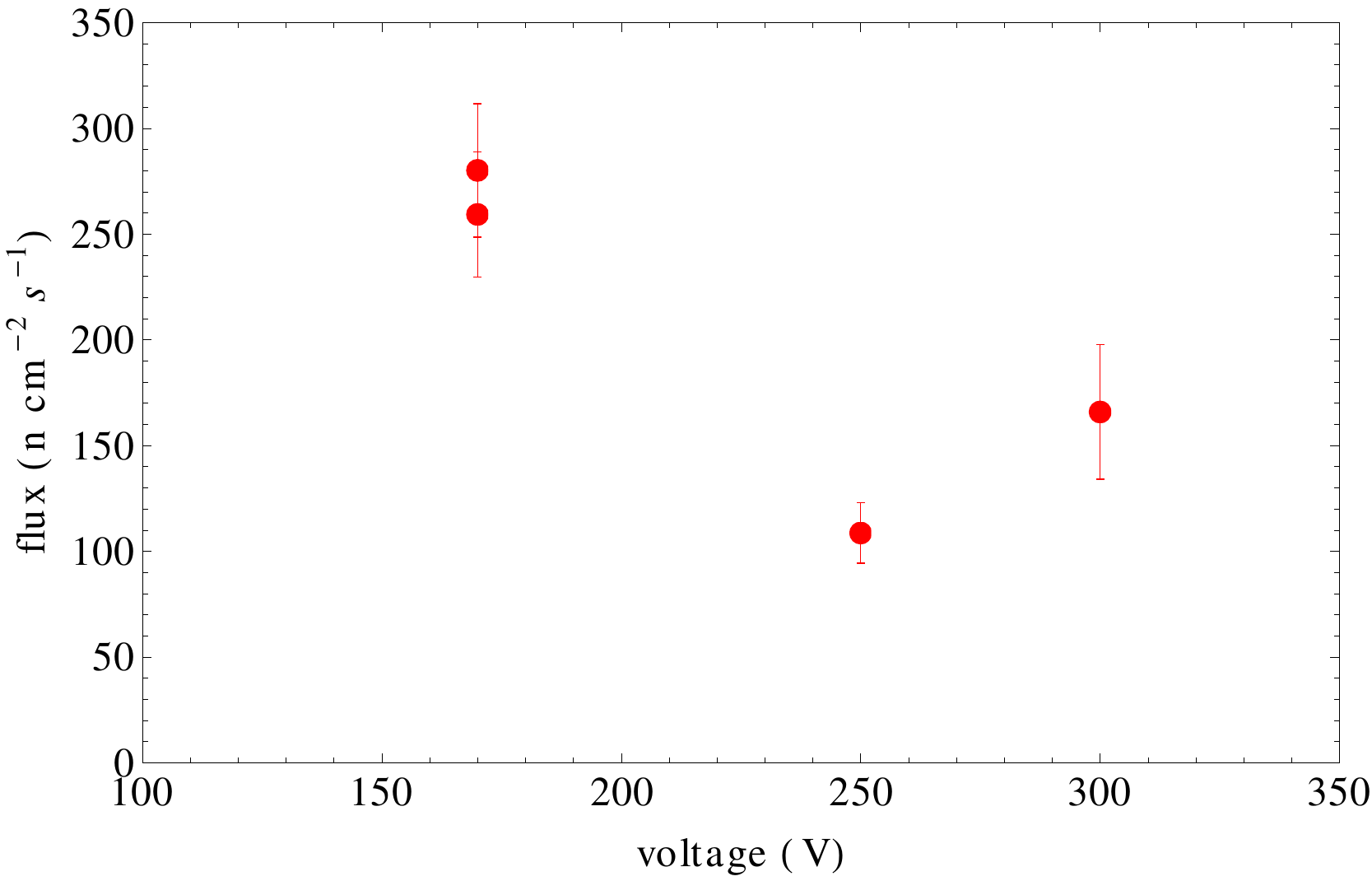}
\caption{Dependence of the neutron flux, as estimated from the track density excess in Run1 data, on the average voltage applied to the electrolytic cell.  }
\label{fig:voltage}
\end{figure}

As shown in Tab.~\ref{tab:results} and Fig.~\ref{fig:results}, the detectors presented a small but significant deviation from the background. Such deviation  did not show 
any dependence on whether the detector was placed inside or outside the cell, nor on the voltage (see Fig.~\ref{fig:voltage}). Even attributing this excess to neutrons produced in the cell, the measured fluxes range between 110 and 280 neutrons~cm$^{-2}$~s$^{-1}$, incompatibly with the $\approx 72000$ neutrons~cm$^{-2}$~s$^{-1}$ reported in Ref.~\cite{Cirillo}.

Indium disks instead never measured any signal, establishing that the thermal (fast) neutron flux is smaller than 1.5 (64) neutrons~cm$^{-2}$~s$^{-1}$ at 95\% C.L.

{\bf{CR-39 background studies}}
\label{sec:bkgd}
The inconsistency between the results obtained with the CR-39 and the Indium techniques led us to further investigate the background subtraction in the analysis of the CR-39 detectors. In the Run1 analysis the background was treated differently from the signal because it had not been wrapped with boron and it was analyzed 40 days before the exposed detectors.

To this aim, in the Run2 and Run3 campaigns  the background detectors were covered with the thin boron film and analyzed simultaneously with the exposed detectors. Furthermore, we studied the time evolution of the background by measuring it at a distance of 60 days.  Fig.~\ref{fig:results} shows a significant increase of background with time (from Run2 to Run3) that we estimated to be $dD/dt=3.8$ tracks~cm$^{-2}$~day$^{-1}$. 

Furthermore, we have left some detectors without boron. As it can be seen in Fig.~\ref{fig:results}, where they are marked as ``naked'', such detectors behave like the background. This suggests that the background is not due to  thermal neutrons, but rather to cosmic rays (that arrive at a rate of $\approx 850$ particles~cm$^{-2}$~day$^{-1}$) or by natural neutrons, that have a similar rate~\cite{AtmNeu}.
Assuming the same calibration as thermal neutrons, $(8.3\pm 0.8)\times 10^{-3}$ tracks/neutron, the observed neutron flux would be 460 neutrons~cm$^{-2}$~day$^{-1}$. We therefore conclude that such background increases with time at a rate that can easily cause significant systematic errors if not properly accounted for. On this basis, we considered flawed the results of the CR-39 detectors during the Run1 data taking and repeated the experiment.

{\bf{Results of the Run2 and Run3 data taking.}}
\label{sec:results}
We have performed the last two data-taking campaigns with the same setup but separated by two months, paying attention to a consistent treatment of exposed and background detectors. To study the nature of a possible excess, we have wrapped some of the detectors in aluminum (to screen electromagnetic radiation) and Cadmium (to screen thermal neutrons).  The calibration constant used is  $(8.3\pm 0.8)\times 10^{-3}$ tracks/neutrons, see above.

The results of both the CR-39 and the Indium detectors, summarized in Tab.~\ref{tab:results} and Fig.~\ref{fig:results}, show that the exposed detectors yield tracks densities consistent with the background. The only two outliers are those that were wrapped in aluminum during Run3, as though the wrapping could cause some spurious tracks. Considering conservatively  the detector with the largest deviation from the background, we set a limit on the neutron flux $\Phi< 105$ neutrons~cm$^{-2}$~day$^{-1}$ from the CR-39 detectors and $\Phi< 0.6 (26)$ neutrons~cm$^{-2}$~day$^{-1}$  from the Indium disks under the assumption that the neutron spectrum is thermal (PM). 

{\bf{Conclusions.}}
We have searched for neutron production in plasma discharges inside an electrolytic cell, with the attempt of reproducing the results reported in Ref.~\cite{Cirillo} with two different types of detectors, CR-39 tracers, as in the original experiment, and Indium disks. The results found and the techniques used are of relevance also to the experiment discussed in~\cite{fulmini} as the underlying physics of neutron production could be the same.  

Since Indium disks never showed a signal, we conclude that  the produced neutron flux is smaller than 1.5 (64) neutrons~cm$^{-2}$~s$^{-1}$ at 95\% C.L. assuming thermal (PM) neutrons.  From the measurements with CR-39 detectors  in Run2 and Run3, when the background was treated properly, we can exclude fluxes larger than 105 neutrons~cm$^{-2}$~s$^{-1}$ at 95\% C.L. Such limits are at least two orders of magnitude smaller than the measured fluxes in Ref.~\cite{Cirillo} (72000 neutrons~cm$^{-2}$~s$^{-1}$).

The initial inconsistency between the results from Indium disks and CR-39 detectors lead us to study the behavior of the latter in detail. We have first verified, contrarily to Ref.~\cite{Cirillo}, that if the layer of boron in which they are immersed is larger than hundreds of $\mu$m the sensitivity to thermal neutrons is lost. 
Next, we demonstrated that CR-39 detectors, regardless of the presence of Boron, integrate  $dD/dt=3.8$ tracks~cm$^{-2}$~day$^{-1}$, likely due to cosmic radiation. It is therefore critical to pay attention to the treatment of the background, since a delay in analyzing the irradiated detectors with respect to the background ones could lead to false positives.

We confirmed that non-irradiated CR-39 show long tails in the track density~\cite{bedogni2} and we noted a small signal in detectors wrapped in aluminum. Therefore, especially when very low neutron fluxes are expected, the use of CR-39 detectors should be accompanied by careful consideration of  disturbing effects. In addition, each batch of CR-39 detectors needs an individual characterization in terms of background distribution and calibration curve.

{\bf{Acknowledgements.}}
We thank L. Maiani and G. Ruocco for  stimulating discussions and S. Loreti(INMRI-ENEA, Rome, Italy) for the CR-39  calibration .


%
%

\end{document}